\documentclass[prb,twocolumn,showpacs,preprintnumbers,amsmath,amssymb]{revtex4}

\usepackage{graphicx}
\usepackage{dcolumn}
\usepackage{bm}

\begin{document}

\preprint{APS/123-QED}

\title{Anisotropic Electron Spin Lifetime in (In,Ga)As/GaAs (110) Quantum Wells}

\author{L. Schreiber}
\author{D. Duda}
\author{B. Beschoten}
\author{G. G\"untherodt}
\affiliation{{II}. Physikalisches Institut, RWTH Aachen
University, and Virtual Institute for Spin Electronics (ViSel),
Templergraben 55, 52056 Aachen, Germany}

\author{H.-P. Sch\"onherr}
\author{J. Herfort}
 \affiliation{Paul-Drude-Institut f\"ur Festk\"orperelektronik, Hausvogteiplatz 5-7,
10117 Berlin, Germany}

\date{\today}

\begin{abstract}
Anisotropic electron spin lifetimes in strained undoped
(In,Ga)As/GaAs (110) quantum wells of different width and height
are investigated by time-resolved Faraday rotation and
time-resolved transmission and are compared to the (001)
orientation. From the suppression of spin precession as a function
of transverse magnetic fields, the ratio of in-plane to
out-of-plane spin lifetimes is calculated. While the ratio
increases with In concentration in agreement with theory, an
unprecedented high anisotropy of 480 is observed for the broadest
quantum well at the low In concentration, when expressed in terms
of spin relaxation times.
\end{abstract}


\maketitle

During the last years, the manipulation of the electron's spin
degree of freedom for information processing was explored in the
new field of spintronics.\cite{Awschalom02} The most prominent
proposal for a logic spin device by Datta and Das,\cite{Datta90}
the spin field effect transistor, is based on the precession of a
spin in a variable electric field due to the Rashba effect.
Recently, more robust devices have been suggested, which are based
on the manipulation of spin relaxation leading to switchable
randomization of the spin orientation
.\cite{Schliemann03,Hall03,Cartoixa03} Some of these proposals
exploit the unique characteristics of bulk inversion asymmetry in
(110)-oriented zinc-blende semiconductor quantum wells (QWs), for
which a large anisotropy of electron spin relaxation time is
predicted.\cite{Dyakonov86}

The largest spin relaxation anisotropy was found in a narrow
$n$-doped (In,Ga)As/(Al,Ga)As QW grown in the [110] crystal
direction.\cite{Morita05} In these zinc-blende semiconductors QWs,
the spin relaxation time $\tau_S$ is mainly governed by the
D'yakonov-Perel' (DP) spin dephasing mechanism, for which
$\tau_S^{DP}$ of 2D confined electron spins is proportional to
$T^{-1}E_g d^{-2} E_1^{-2}\tau_p^{-1}$, where $T$ is the
temperature, $\tau_p$ the momentum relaxation time, $E_1$ the
quantized kinetic energy of electrons and $E_g$ and $d$ are the
band gap and the thickness of the QW layer, respectively.
\cite{Dyakonov86} Additionally, the effect of DP spin dephasing
strongly depends on the QW's confinement direction as well as on
the spin direction:\cite{Dyakonov86} For (001) QWs, the relaxation
time of spins pointing in the in-plane direction
$\tau^{\parallel}$ and in the out-out-plane direction
$\tau^{\perp}$ is given by $\tau_{S(001)}^{\parallel}=\tau_S^{DP}$
and $\tau_{S(001)}^{\perp}=\tau_S^{DP}/2$, respectively, whereas
it is $\tau_{S(110)}^{\parallel}=4 \tau_S^{DP}$ and
$\tau_{S(110)}^{\perp}=\infty$ for (110) QWs, leading to infinite
spin relaxation anisotropy for the latter. Indeed, large
$\tau_{S(110)}^{\parallel}>$~1~ns as well as spin relaxation
anisotropy were experimentally found in $n$-doped
GaAs/Al$_{0.4}$Ga$_{0.6}$As (110)\cite{Ohno99,Doehrmann04} and in
$n$-doped In$_{0.08}$Ga$_{0.92}$As/Al$_{0.4}$Ga$_{0.6}$As (110)
QWs.\cite{Morita04,Morita05} Additional isotropic spin relaxation
mechanisms have to be taken into account: e.g., the
Bir-Aronov-Pikus (BAP) mechanism \cite{Bir75} based on randomly
oriented hole spins acting on electron spins via exchange
interaction is believed to be dominant only at low temperature in
undoped and $n$-doped QWs.\cite{Doehrmann04} However, there has
been no detailed study of anisotropic spin lifetimes in shallow
and strained In$_x$Ga$_{1-x}$As/GaAs (110) QWs yet.

In the present work, we therefore investigated spin precession as
a function of transverse magnetic fields of three
In$_x$Ga$_{1-x}$As/GaAs (110) QWs with different QW width and In
concentration $x$ employing the time-resolved Faraday rotation
(TRFR) technique. The QWs were not intentionally doped, allowing
us to separate the magnetization loss due to electron spin
relaxation in the conduction band characterized by the spin
relaxation time $\tau_S^{\perp,\parallel}$ and the carrier
recombination processes described by the carrier lifetime
$\tau_R$. The decay of electron spin magnetization as observed by
TRFR is then given by the spin lifetime $T_S^{\perp,\parallel}$
\cite{OrientationKap2, comment2}
\begin{equation}\label{relaxation time}
 \left(T_S^{\perp,\parallel}\right)^{-1}=\left(\tau_S^{\perp,\parallel}\right)^{-1}+\tau_R^{-1},
\end{equation}
when hole spins are considered to be already relaxed. Carrier
lifetimes were measured using time-resolved transmission (TRTR)
simultaneously to TRFR, in order to determine electron spin
relaxation times from the measured spin lifetimes. All
measurements were performed using a picosecond-pulsed laser with
energies tuned to the lowest QW electron-hole transition confirmed
independently by TRTR and photoluminescence (PL) at each
temperature. The small spectral width of the picosecond laser
(FWHM 0.5~nm) reduces the energetical width of occupied states and
thus lowers the effect of inhomogeneous dephasing on
$\tau_S^\perp$,\cite{Kikkawa98} which otherwise may mask the
anisotropy caused by the DP mechanism. In this way, investigating
our broadest In$_x$Ga$_{1-x}$As/GaAs (110) QW, we observed the
highest spin relaxation anisotropy ever reported. We observed an
increase of anisotropy with the increase of In-concentration which
is consistent with DP and confirmed that the anisotropy does not
vanish at 10~K as has been previously observed for a $n$-doped
(In,Ga)As/(Al,Ga)As (110) QW.\cite{Morita05}

For all samples, the QW is formed by a single undoped and strained
In$_x$Ga$_{1-x}$As-layer sandwiched between intrinsic GaAs
barriers grown by molecular-beam epitaxy. Three samples with
different In$_x$Ga$_{1-x}$As-layer thickness $d$ and
In-concentration $x$ were grown on (110)-oriented semi-insulating
GaAs substrates: samples A and B have the same $d =$~8~nm and In
concentrations of $x =$~11.5~\% and $x =$~19~\%, respectively.
Sample C has a much broader QW layer of $d =$~14~nm, but the same
In concentration as sample A. Reference sample D was grown under
the same conditions as sample A, but on a (001)-oriented GaAs
substrate. For simultaneous TRFR and TRTR measurements we employed
a mode-locked Ti:Al$_2$O$_3$ laser generating 1~ps optical pulses
with 80~MHz repetition frequency. Normal-incident pump pulses,
which were alternatingly left- and right-circularly polarized by a
photo-elastic modulator (PEM) working at 42~kHz and chopped at
1600~Hz frequency, excite spin-polarized electrons and holes along
the growth direction. The evolution of the corresponding
magnetization projected along this direction was determined by
means of the Faraday rotation angle $\theta_F$ of transmitted
linear-polarized probe pulses, which were delayed by a variable
time $\Delta t$ with respect to the pump pulses. A balanced
diode-bridge was employed to measure $\theta_F$ and the output was
locked to the PEM frequency. The evolution of the filling of the
QW is determined by the change of transmission $\Delta
\zeta(\Delta t)$ of the probe pulse and recorded by the sum of the
intensities on the diode bridge locked to the chopper frequency of
the pump beam.

\begin{figure}
\includegraphics{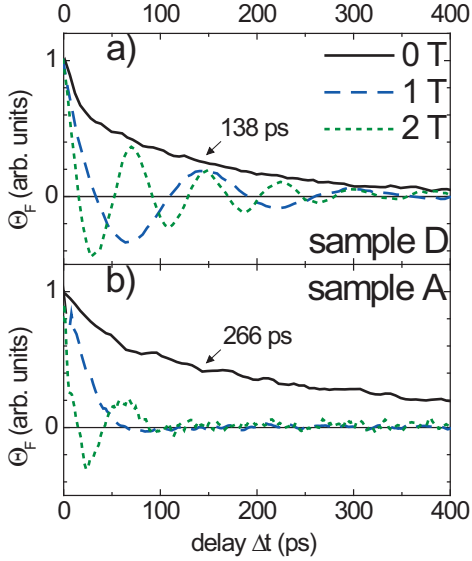}
\caption{\label{fig:epsart} (Color online) Faraday rotation angle
$\theta_F(\Delta t)$ of (001)-oriented sample D (a) and
(110)-oriented sample A (b) for various transverse magnetic fields
$B_{ext}$ measured at 70 K.}
\end{figure}

Figure 1 displays $\theta_F(\Delta t)$ for (001)-oriented sample D
and (110)-oriented sample A at various transverse magnetic fields
$B_{ext}$. Data were taken at 70~K, at which the longest electron
spin lifetime as a function of temperature was found for sample A.
Fitting the curves for zero magnetic field starting from $\Delta t
=$~50~ps reveal a single-exponential decay reflecting the electron
spin lifetime $T_S^{\perp}$ to be 138~ps and 266~ps for samples D
and A, respectively. An additional fast decay ($<$~10~ps at 70~K)
might be caused by hole spin relaxation.\cite{comment1} In a
transverse magnetic field $B_{ext}\neq 0$, spin precession is
observed in the control sample D for all $B_{ext}$, whereas spin
precession is not observed at $B_{ext}\lesssim$~1~T for the
(110)-oriented sample A. Additionally, the electron spin lifetime
of sample A is dramatically reduced when a transverse magnetic
field is applied similar to Refs. \cite{Morita04, Morita05}. The
corresponding carrier lifetimes $\tau_R$ at 70 K (not shown)
turned out to be independent of the applied $B_{ext}$ and are
600~ps and 675~ps for sample D and A, respectively, which reflects
that $\tau_R$ does not depend on the electron spin direction.

In order to model the complex $B_{ext}$-dependence for the
anisotropic sample A, the magnetization of electrons spins normal
($S_{\perp}$) and parallel ($S_{\parallel}$) to the QW plane has
to be considered separately. The time evolution of $\vec{S}$ in an
in-plane magnetic field is given by \cite{Doehrmann04}
\begin{equation}\label{diffequation}
    \frac{\partial}{\partial t} \left(\begin{array}{c} S_{\parallel} \\ S_{\perp} \\ \end{array} \right)
    = - \left( \begin{array}{cc} \Gamma_{\parallel} & -\omega \\ \omega & \Gamma_{\perp} \end{array} \right)
    \left(\begin{array}{c} S_{\parallel} \\ S_{\perp} \\ \end{array}
    \right),
\end{equation}

\noindent where $\omega=g \mu_B B_{ext}/\hbar$ is the Larmor
frequency. For undoped QWs the total relaxation rates
$\Gamma_{\perp,\parallel}=(T_S^{\perp, \parallel})^{-1}$ are the
sum of the anisotropic electron spin relaxation rates
$\gamma_{\perp,\parallel}=(\tau_S^{\perp,\parallel})^{-1}$ and the
isotropic carrier recombination rate $(\tau_R)^{-1}$ in the QW:
$\Gamma_{\perp,\parallel}= \gamma_{\perp,\parallel}+(\tau_R)^{-1}$
in accordance with Eq. \ref{relaxation time}. The solution for
$S_{\perp}$, which is measured by TRFR, is given by

\begin{equation}\label{perpmagnetization}
 S_{\perp}(t)=\frac{S_0}{\cos(\phi)}\exp\left(- \frac{\Gamma_{\perp}+\Gamma_{\parallel}}{2} t
 \right)\cos(\omega^{\prime} t - \phi),
\end{equation}

\noindent with $\tan(\phi)=(\Gamma_{\perp}-\Gamma_{\parallel})/(2
\omega^{\prime})$ and the modified Larmor frequency

\begin{equation}\label{Larmormod}
  \omega^{\prime} =
  \sqrt{\omega^2-\left(\frac{\Gamma_{\parallel}-\Gamma_{\perp}}{2}\right)^2}.
\end{equation}

\begin{figure}
\includegraphics{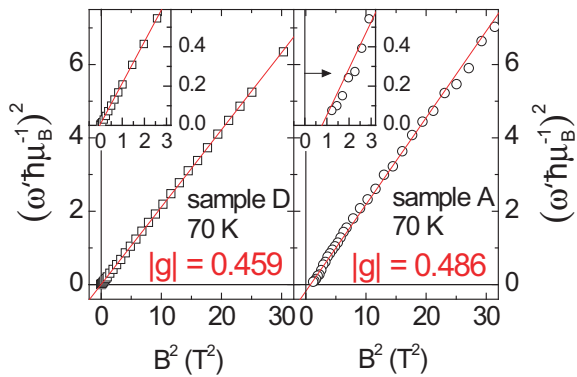}
\caption{\label{fig:epsart} (Color online) Square of measured
Larmor frequency $\omega^{\prime2}(B^2_{ext})$ for (001)-oriented
sample D (squares) and (110)-oriented sample A (circles) measured
at 70 K; solid lines are linear fits; zoom near $B=$~0~T in the
insets.}
\end{figure}

In order to calculate the spin lifetime anisotropy, the square of
the measured modified Larmor frequency $\omega^{\prime 2}$ of
sample A and D is plotted as a function of $B_{ext}^2$ as
displayed in Figure 2. In the inset, a right-shift of the plot for
sample A is obvious and indicates the suppression of precession in
the (110)-oriented QW. According to Eq.~\ref{Larmormod} this shift
is caused by the difference of the total spin relaxation rates
$\Gamma_{\parallel}-\Gamma_{\perp}$ and thus by the spin lifetime
anisotropy. From the right-shift and the slope of the linear fit,
both the difference of the relaxation rates
$\Gamma_{\parallel}-\Gamma_{\perp}$ and the absolute electron
g-factor can be determined, respectively. Using
$\Gamma_{\parallel}-\Gamma_{\perp}$ and assuming
$T_S^{\perp}=$~266~ps to be independent of $B_{ext}$
\cite{Morita05}, the in-plane spin lifetime of sample A is
calculated to be $T_S^{\parallel} =$ (13~$\pm$~2)~ps. Thus, the
anisotropy of the spin lifetime $T_S^{\perp}/T_S^{\parallel}$
turns out to be 21~$\pm$~2 for sample A. The last assumption is
justified, since the exponential decay in
Eq.~\ref{perpmagnetization} is observed to be independent of the
magnetic field in the oscillatory regime and simulations using
constant $T_S^{\perp}$ fit well in the non-oscillatory regime,
which is demonstrated here for sample C in Figure 3a.

\begin{figure}
\includegraphics{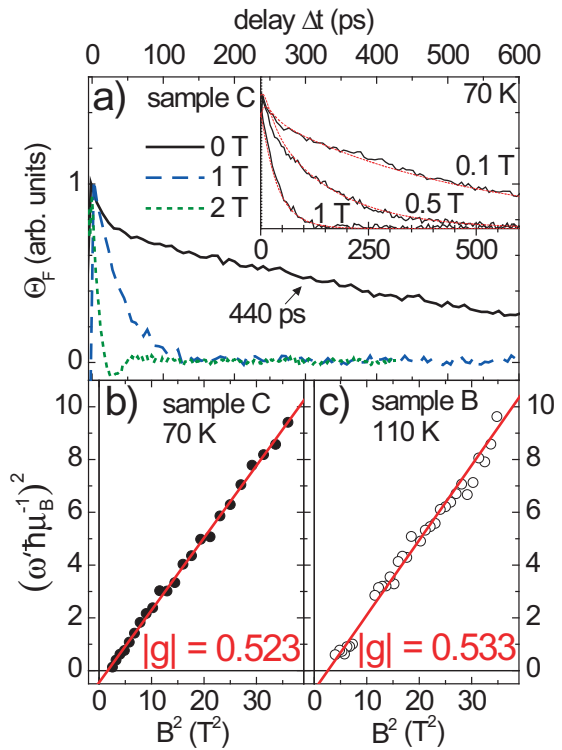}
\caption{\label{fig:epsart} (Color online) (a) Faraday rotation
angle $\theta_F(\Delta t)$ for the broad (110) QW C at various
transverse magnetic fields $B^2_{ext}$; inset: measurements (line)
and simulations (dotted line) for imaginary $\omega^{\prime}$;
Larmor frequency $\omega^{\prime2}(B_{ext})$ for (b) (110) QW C
and (c) (110) QW B measured at 70~K and 110~K, respectively; solid
lines are linear fits.}
\end{figure}

\begin{figure}
\includegraphics{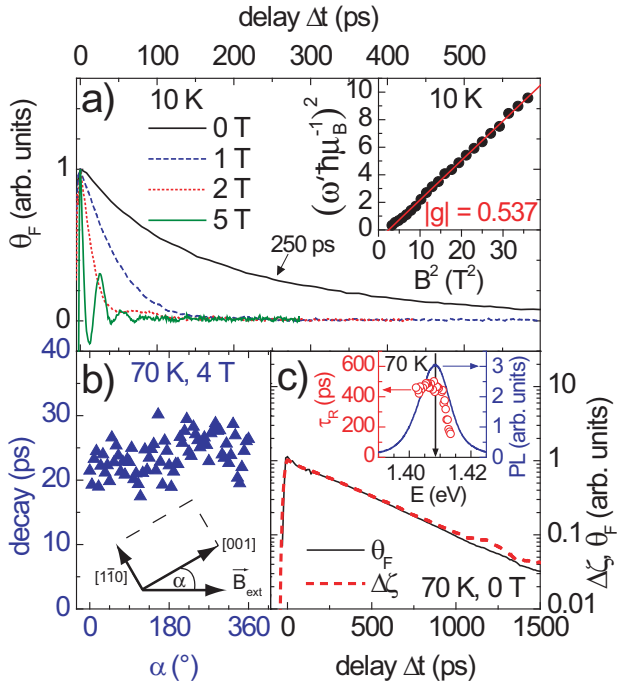}
\caption{\label{fig:epsart} (Color) (a) Faraday rotation angle
$\theta_F(\Delta t)$ and fitted Larmor frequency
$\omega^{\prime2}(B_{ext})$ (inset) for the broad (110) QW C at
various transverse magnetic fields $B_{ext}$ measured at 10~K; (b)
fitted decay of $\theta_F(\Delta t)$ measured as a function of
angle $\alpha$ between $B_{ext}$ and [001] crystal direction at
4~T and 70~K for sample C; (c) comparison of $\theta_F(\Delta t)$
and $\Delta \zeta(\Delta t)$ for sample C at 0~T and 70~K measured
with laser energy $E$ as marked in the inset, where the PL
spectrum is compared to $\tau_R(E)$.}

\end{figure}

In the following, this result is compared to the other
(110)-oriented QWs. Figure 3 shows $\theta_F(\Delta t)$ at various
$B_{ext}$ for sample C as well as $\omega^{\prime 2}(B_{ext}^2)$
for sample C and B. Data were taken at 110~K and 70~K for sample B
and sample C, respectively, at which the maximum of
$T_S^{\perp}=$~335~ps and $T_S^{\perp}=$~440~ps was measured.
Comparing $\theta_F(\Delta t)$ of sample C to sample A (see Fig.
1b), $T_S^{\perp}$ of sample C is found to be longer, but spin
precession turns out to be even more suppressed, which indicates a
higher anisotropy. The same can be concluded from the pronounced
right-shift of $\omega^{\prime 2}(B_{ext}^2)$ (see Fig. 3b). Using
the derivation explained above, the anisotropy of the spin
lifetime $T_S^{\perp}/T_S^{\parallel}$ follows to be 52~$\pm$~10
and 54~$\pm$~6 with $T_S^{\parallel}=$ (6.5~$\pm$~0.8)~ps and
$T_S^{\parallel}=$ (8.2~$\pm$~0.5)~ps for sample B and C,
respectively. Compared to the calculated value of sample A, the
result for the deeper QW B has the correct tendency, since DP is
more dominant for the latter according to theory. However, the
high anisotropy for the broadest QW C is surprising and needs
further discussion. The effectiveness of DP should be reduced for
sample C, since $\tau_S^{DP} \propto d^{-2}E_1^{-2} \propto d^2$.
The increased spin lifetime as well as the higher anisotropy
compared to sample A might than be explained by a reduction of an
isotropic spin relaxation channel. Scattering at the barrier
interfaces, which has not been considered yet, might be a suitable
candidate, since it can be assumed to be dependent on the
volume-to-interface ratio and thus less pronounced for the
broadest QW C.

We now check the consistency of our results with our simple model
and with DP theory by varying the temperature and the direction of
the magnetic field. Since for the derivation of
$\Gamma_{\parallel}-\Gamma_{\perp}$ and the $g$ factor only the
oscillatory curves are considered as explained above, consistency
of raw data with Eq. \ref{perpmagnetization} is checked for
imaginary $\omega^{\prime}$. With only the amplitude $S_0$ as a
free parameter left, the simulations fit well to the measured
$\theta_F(\Delta t)$ curves of sample C, as can be seen from the
inset of Figure 3. Furthermore, we measured the anisotropy of
sample C at 10 K, at which the isotropic BAP mechanism is
considered to be dominant and thus the anisotropy of spin
lifetimes should be reduced. Indeed, a lower anisotropy of
$T_S^{\perp}/T_S^{\parallel}=$~39~$\pm$~9 is found for sample C at
10~K.\cite{comment1} Since the anisotropy does not vanish, DP is
still present at 10~K as was already observed for an
(In,Ga)As/(Al,Ga)As (110) QW.\cite{Morita05} Finally, we checked
for any in-plane anisotropy of the spin lifetime by rotating the
sample in-situ around the growth direction in a transverse
magnetic field. Figure 4 (b) displays no systematic changes of the
decay for sample C at 70~K in the oscillatory ($B_{ext} =$~4~T)
regime. Thus, in consistency with theory
\cite{Dyakonov86,Wayne02}, spin relaxation is equivalent for all
spin directions in the plane of the QW.

In the end, we express the calculated anisotropy of spin lifetimes
in terms of spin relaxation times using the carrier lifetime
$\tau_R$. Since $\Delta \zeta(\Delta t)$ is found to be
independent of the transverse magnetic field (not shown) and thus
$\tau_R$ does not change with the spin direction, we can use Eq.
\ref{relaxation time} for our undoped QWs. The difference of the
in-plane and out-of-plane electron spin relaxation rates
$\gamma_{\parallel}-\gamma_{\perp}=\Gamma_{\parallel}-\Gamma_{\perp}$
can be easily seen from the right-shift of the linearly fitted
$\omega^{\prime 2}(B_{ext}^2)$ leading to the anisotropy of spin
relaxation times. For sample C, $\Delta \zeta(\Delta t)$ and
$\theta_F(\Delta t)$ are measured simultaneously at zero magnetic
field, 70~K and with a laser energy $E = 1.409$~eV as displayed in
Figure 4c. Fitting $\Delta \zeta(\Delta t)$ and $\theta_F(\Delta
t)$, we determine $\tau_R=$ (450~$\pm$~20)~ps and $T_S^{\perp}=$
(440~$\pm$~20)~ps leading to $\tau_S^{\perp}>$~4~ns. Data were
taken in an energy region of the PL peak of the QW, where both
$\tau_R$ and $T_S$ are found to be constant (see inset of Figure
4c). Since $\tau_R \gg T_S^{\parallel}$, the limitation of the
carrier lifetime for the calculation of $\tau_S^{\parallel}$ is
negligible according to Eq. \ref{relaxation time}. Expressing the
anisotropy in terms of spin relaxation times, we thus obtain
$\tau_S^{\perp}/\tau_S^{\parallel}>480$ for sample C at 70~K. This
lower limit of the anisotropy is by a factor of eight higher than
the maximum anisotropy observed so far.\cite{Morita05} As the
carrier lifetime $\tau_R =$~675~ps of sample A is considerably
higher than its out-of-plane spin lifetime $T_S^{\perp} =$~266~ps
there is not much difference found when the anisotropy is
expressed instead of spin lifetime in terms of spin relaxation
time: $\tau_S^{\perp}/\tau_S^{\parallel}=$ 34~$\pm$~7.

In conclusion, we observed anisotropic electron spin lifetimes in
strained and undoped InGaAs/GaAs (110) QWs by the suppression of
spin precession. Increasing the In concentration enhances the
dominating DP mechanism and leads to a higher anisotropy
consistent with theory. The broadest QW exhibit longest
$T_S^{\perp}$ as well as an unprecedented large anisotropy, which
in terms of spin relaxation times exceeds 480 and does not vanish
at low temperatures, when the isotropic BAP is most relevant. No
in-plane anisotropy was found and the suppression of spin
precession can be described by $T_S^{\perp},T_S^{\parallel}$ and
the electron g-factor in a transverse magnetic field.

We gratefully acknowledge Dr. R. Hey (PDI) for discussion about
(110) growth. The work was supported by BMBF / FKZ 13N8244 and by
HGF.


\begin{thebibliography}{15}
\bibitem[{com()}]{comment1}
\bibinfo{note}{Curve fitting is more complex at low temperatures, since the
  additional decay time in $\theta_F(\Delta t)$ increases with the decrease of the temperature. However, this decay is too short to determine its anisotropy.}
\end{thebibliography}
\end{document}